\documentclass[11pt,a4paper]{article}
\usepackage{jheppub}
\usepackage[utf8]{inputenc}
\usepackage{amsmath,amsfonts}
\usepackage{enumitem}
\usepackage{caption}
\usepackage{subcaption}
\usepackage{graphicx}
\usepackage{slashed}
\usepackage{booktabs}
\usepackage{mciteplus}
\usepackage[dvipsnames]{xcolor}
\allowdisplaybreaks

\def\eqn#1{eq.~\eqref{#1}}
\def\eqns#1#2{eqs.~\eqref{#1} and~\eqref{#2}}

\def\be{\begin{equation}}
\def\ee{\end{equation}}
\def\bea{\begin{eqnarray}}
\def\eea{\end{eqnarray}}
\def\beal{\begin{equation}\begin{aligned}}
\def\eeal{\end{aligned}\end{equation}}
\def\nn{\nonumber}

\def\cM{\mathcal{M}}
\def\cN{\mathcal{N}}
\def\cO{\mathcal{O}}

\def\cR{\mathcal{R}}

\def\Tr{\text{Tr}}
\def\eps{\epsilon}

\usepackage{stackengine}

\preprint{UUIPT-59/21}

\title{The gluon Regge trajectory at three loops from planar Yang-Mills theory}

\author[a,b,1]{Vittorio Del Duca}
\author[c]{Robin Marzucca}
\author[d]{Bram Verbeek}
\note{On leave from INFN, Laboratori Nazionali di Frascati, Italy.}
\affiliation[a]{Institute for Theoretical Physics, ETH Z\"urich, 8093 Z\"urich, Switzerland.}
\affiliation[b]{Physik-Institut, Universit\"at Z\"urich, Winterthurerstrasse 190, 8057 Z\"{u}rich, Switzerland}
\affiliation[c]{Niels Bohr Institute, Copenhagen University, Blegdamsvej 17, 2100 Copenhagen \O , Denmark}
\affiliation[d]{Department of Physics and Astronomy, Uppsala University, 75108 Uppsala, Sweden}

\emailAdd{delducav@itp.phys.ethz.ch}
\emailAdd{robin.marzucca@nbi.ku.dk}
\emailAdd{bram.verbeek@physics.uu.se}

\abstract{We compute the three-loop leading-colour corrections to the Yang-Mills Regge trajectory and gluon impact factor. 
Conjecturing that, in analogy with ${\cal N}=4$ super Yang-Mills (SYM), in a suitable scheme $N_c$-subleading terms are absent from the three-loop Regge trajectory, we understand our result as the first computation of the pure gauge, or $n_f = 0$, part of the QCD three-loop Regge trajectory.
The results are presented both for the bare and renormalised amplitudes and are consistent with predictions from infrared factorisation along with reproducing known results in planar $\mathcal{N}=4$ SYM through a maximal weight truncation.
We also include the dependence on a Regge factorisation scale to facilitate future applications in BFKL theory at next-to-next-to leading logarithmic accuracy.}


\begin{document}

\maketitle


\section{Introduction}
\label{sec:intro}

The Regge limit of scattering amplitudes has been the subject of many theoretical as well as phenomenological studies. As $s/|t| \rightarrow \infty$, scattering amplitudes exhibit logarithmic divergences which can be resummed to all orders at leading logarithmic (LL)~\cite{Lipatov:1976zz,Kuraev:1976ge,Kuraev:1977fs,Balitsky:1978ic} and next-to-leading logarithmic (NLL) accuracy~\cite{Fadin:1998py,Ciafaloni:1998gs,Kotikov:2000pm,Kotikov:2002ab} in $\log(s/|t|)$. This is of great phenomenological interest, as fixed-order calculations in the coupling constant will generally diverge in the forward-scattering limit, which makes it difficult to predict the behaviour of scattered particles close to the beam line. Resumming the large logarithms which appear in this limit allows one to improve the description of the inter-jet activity in events at the Large Hadron Collider (LHC) where large rapidity intervals are spanned~\cite{Mueller:1986ey,DelDuca:1993mn,Stirling:1994he,Andersen:2001kta,Colferai:2010wu,Andersen:2011hs,Ducloue:2013hia}. Furthermore, in the last decade the study of the Regge limit has deepened after realising that it is a useful kinematic constraint for QCD and ${\cal N}=4$ super Yang-Mills (SYM) amplitudes in 
bulk kinematics, and that in the Regge limit
amplitudes in ${\cal N}=4$ SYM~\cite{Dixon:2012yy,DelDuca:2016lad,DelDuca:2019tur}
and amplitudes~\cite{Caron-Huot:2020grv} and cross sections~\cite{DelDuca:2013lma,DelDuca:2017peo} in QCD are endowed with a rich mathematical structure.

In the Regge limit, the squared centre-of-mass energy $s$ is much larger than the momentum transfer $t$, $s\gg |t|$, and
$2\to 2$ scattering amplitudes are dominated by gluon exchange in the $t$ channel. Contributions which do not feature gluon exchange in the $t$ channel
are power suppressed in $t/s$. At tree level we can write the $2\to 2$ amplitudes in a factorised way. For example, 
the tree amplitude for gluon-gluon scattering $g_1\, g_2\to g_3\,g_4$ may be written as \cite{Lipatov:1976zz,Kuraev:1976ge}, 
\begin{equation}
\cM^{(0)}_{4g} = 
    \left[g_s (F^{a_3})_{a_2c}\, C^{g(0)}(p_2, p_3) \right]
    {s\over t} \left[g_s (F^{a_4})_{a_1c}\, C^{g(0)}(p_1, p_4) \right]\, ,\label{elas}
\end{equation}
with $s=(p_1+p_2)^2$, $q = p_2 + p_3$ and $t=q^2  \simeq -|q_\perp|^2$, and $(F^c)_{ab} = i\sqrt{2} f^{acb}$.
The tree level impact factors, $C^{g(0)}$, depend on the momenta and the helicities of the outgoing gluons~\cite{DelDuca:1995zy}, however
in what follows their precise definition is immaterial since they will be factored out.
As it is apparent from the colour coefficient  $(F^{a_3})_{a_2c} (F^{a_4})_{a_1c}$, in \eqn{elas} only the antisymmetric octet ${\bf 8}_a$ is exchanged in the $t$ channel.

Since $u=-s-t\simeq -s$ in the Regge limit, $2\to 2$ scattering amplitudes are symmetric under $s\leftrightarrow u$ crossing,
and we may consider amplitude states whose kinematic and colour coefficients have a definite signature under $s\leftrightarrow u$ crossing,
\begin{equation}
    \cM_4^{(\pm)}(s,t) = \frac{ \cM_4(s,t) \pm \cM_4(u,t) }{2}\,,
\label{eq:sucross}
\end{equation}
such that $\cM_4^{(-)}(s,t)$ ($\cM_4^{(+)}(s,t)$) has kinematic and colour coefficients which are both odd (even) 
under $s\leftrightarrow u$ crossing.
Higher-order contributions to $g g \rightarrow g g$ scattering
in general involve additional colour structures, as dictated by the decomposition of the product ${\bf 8}_a \otimes {\bf 8}_a$
into irreducible representations,
\be
  {\bf 8}_a \otimes {\bf 8}_a \, = \, \{ {\bf 1} \oplus {\bf 8}_s \oplus {\bf 27} \} 
  \oplus [ {\bf 8}_a \oplus {\bf 10} \oplus \overline{{\bf 10}} ]\ \, ,
\label{8*8}
\ee
where the curly (square) brackets collect the representations which are even (odd) under $s\leftrightarrow u$ crossing.
In what follows, we shall consider only the antisymmetric octet ${\bf 8}_a$, and we shall drop the $(-)$ parity label 
under $s\leftrightarrow u$ crossing.

When loop corrections to the tree amplitude (\ref{elas}) are considered, we write the four-gluon amplitude for the
antisymmetric octet ${\bf 8}_a$ exchange in the $t$ channel as
\begin{equation}\label{eq:factorisation&remainder}
	\cM^{[8_a]}_{4g} = \cM^{[8_a]\,\text{fact}}_{4g} +   \cR^{[8_a]}_{4g} \,,
\end{equation}
where $\cM^{[8_a]\,\text{fact}}_{4g}$ is the part of the amplitude which factorises and $\cR^{[8_a]}_{4g}$ denotes the non-factorising part often referred to as the remainder. The factorised part can be written as~\cite{Fadin:1993wh},
\begin{equation}\label{eq:ReggeFactorisation}
	\cM^{[8_{a}]\, \text{fact}}_{4g} = \frac{s}{t} \left[g_s (F^{a_3})_{a_2c}\, C^{g}(p_2, p_3) \right]
	\left[ \left( \frac{-s}{\tau} \right)^{\alpha(t)} + \left( \frac{s}{\tau} \right)^{\alpha(t)} \right] 
	\left[g_s (F^{a_4})_{a_1c}\, C^{g}(p_1, p_4) \right]
	 \,,
\end{equation}
where $\tau > 0$ is a Regge factorisation scale,
which is of order of $t$ (and thus much smaller than $s$)\footnote{The precise definition of $\tau$ is immaterial to LL accuracy, where one can suitably fix $\tau = - t$.} and where
the impact factors,
\be \label{eq:permExpCGlu}
	C^g(t) = C^{g(0)}(t) \left( 1 + \sum_{L=1}^{\infty}N_c^L \tilde{g}_s ^{2L} C^{g(L)}(\epsilon) \right) \,,
\ee
and the Regge trajectory,
\be \label{eq:permExpTraj}
\alpha(t) = \sum_{L=1}^{\infty} N_c^L  \tilde{g}_s ^{2L}  \alpha^{(L)}(\epsilon)\,,
\ee
with $N_c$ the number of colours,
are expanded in the strong coupling $\alpha_s = g_s^2/(4\pi)$ through the rescaled coupling,
\begin{equation}\label{eq:gtildedef}
	\tilde{g}_s^2 = \frac{\alpha_s}{4 \pi} \frac{\kappa_{\Gamma}}{S_\epsilon} \left(\frac{\mu^2}{-t}\right)^{\epsilon} \,,
\end{equation}
with
\begin{equation}
	\kappa_{\Gamma} = (4\pi)^\epsilon \frac{\Gamma(1+\epsilon)\Gamma^2(1-\epsilon)}{\Gamma(1-2\epsilon)}\,, \qquad S_\epsilon = (4 \pi)^{\epsilon} e^{-\epsilon \gamma_E}\,,
\end{equation}
where $\gamma_E$ denotes the Euler-Mascheroni constant.
We can write the amplitude (\ref{eq:factorisation&remainder}) as a double expansion in the rescaled coupling $\tilde{g}_s$
and in $\log(s/\tau)$, 
\begin{align}
\label{eq:OctetPerturbExpansion}
	\cM^{[8_a]}_{4g} &=
	\cM^{[8_a](0)}_{4g}\left(1 + \sum_{L=1}^{\infty} N_c^L \tilde{g}_s ^{2L}
	\cM^{[8_a] (L)}_{4g} \right) \\
	 &= \cM^{[8_a](0)}_{4g}\left(1 + \sum_{L=1}^{\infty} N_c^L \tilde{g}_s ^{2L} \sum_{i=0}^{L}  \cM^{(L,i)}_{4g} \log^i\left( \frac{s}{\tau} \right)  \right)\,,\label{eq:ReggeLogExpansion}
\end{align}
where we factored out the tree amplitude $\cM^{[8_a](0)}_{4g}$, and where
the $M^{(L,L)[8_a]}_{4g}$ coefficients are referred to as having LL accuracy, the $M^{(L,L-1)[8_a]}_{4g}$ coefficients have NLL accuracy,
and in general the $M^{(L,L-k)[8_a]}_{4g}$ coefficients have $\mathrm{N^kLL}$ accuracy. The remainder term in \eqn{eq:factorisation&remainder},
\begin{equation}\label{eq:ReggeLogExpansionRem}
	\cR^{[8_a]}_{4g} =\sum_{L=2}^{\infty} N_c^L \tilde{g}_s^{2L} \sum_{i=0}^{L-2}  \cR^{(L,i)}_{4g} \log^i\left( \frac{s}{\tau} \right) \,.
\end{equation}
occurs first at next-to-next-to-leading logarithmic (NNLL)
accuracy through the non-logarithmic term, $\cR^{(2,0)}_{4g}$.

At LL accuracy in $\log(s/|t|)$, the four-gluon amplitude is real, and the antisymmetric octet ${\bf 8}_a$
is the only colour representation exchanged in the $t$ channel
to all orders in $\alpha_s$~\cite{Lipatov:1976zz,Kuraev:1976ge},
\begin{equation}
\cM_{4g}\big|_{LL} =  \cM^{[8_a]\, \text{fact}}_{4g}\big|_{LL} \,.
\end{equation}
The exponentiation of $\log(s/\tau)$ in \eqn{eq:ReggeFactorisation} is called gluon Reggeisation, and we say that in the Regge limit the four-gluon amplitude (\ref{eq:ReggeFactorisation}) features the exchange in the $t$ channel of one Reggeised gluon, or Reggeon.
The one-loop amplitude at LL accuracy allows one to determine the one-loop Regge trajectory,
\be
 \cM^{(1,1)}_{4g} = \alpha^{(1)}\,.\label{eq:oneloopLL} 
\ee
Here and henceforth, for the sake of brevity, we drop the explicit $\epsilon$-dependence.

At NLL accuracy, the real part of the amplitude is still given only by the antisymmetric octet ${\bf 8}_a$, through \eqn{eq:ReggeFactorisation}~\cite{Fadin:1993wh},
\begin{equation}
 {\rm Re}\left[ \cM_{4g} \right]_{NLL} = {\rm Re}\left[ \cM^{[8_a]\, \text{fact}}_{4g} \right]_{NLL} \,, 
\end{equation}
i.e. gluon Reggeisation holds at NLL accuracy~\cite{Fadin:2006bj,Fadin:2015zea}.
The real parts of the one-loop and of the two-loop amplitudes
allow one to determine the one-loop impact factor~\cite{Fadin:1993wh,Fadin:1992zt,Fadin:1993qb,DelDuca:1998kx,Bern:1998sc} and the two-loop Regge trajectory~\cite{Fadin:1995xg,Fadin:1996tb,Fadin:1995km,Blumlein:1998ib,DelDuca:2001gu}, respectively,
\begin{align}
	{\rm Re}\left[ \cM^{(1,0)}_{4g}  \right] &= 2 C^{g(1)}\,.\label{eq:oneloopNLL} \\
	{\rm Re}\left[ \cM^{(2,1)}_{4g}  \right] &= \alpha^{(2)} + 2 C^{g(1)}\alpha^{(1)}\,.\label{eq:twoloopNLL}
\end{align}

At NNLL accuracy, 
a three-Reggeon exchange~\cite{Fadin:2016wso,Caron-Huot:2017fxr,Fadin:2017nka,Caron-Huot:2020grv} occurs in the four-gluon amplitude for the antisymmetric octet ${\bf 8}_a$.
It reveals itself in the non-logarithmic term of the two-loop amplitude~\cite{DelDuca:2001gu} as a violation of the universality of Regge factorisation between different parton flavours implied by \eqn{eq:ReggeFactorisation}, which can also be analysed through infrared factorisation~\cite{Bret:2011xm,DelDuca:2011ae,DelDuca:2013ara,DelDuca:2014cya}. We parametrise the violation of factorisation through the remainder term $\cR^{[8_a]}_{4g}$ of \eqn{eq:factorisation&remainder}.
The two-loop impact factor $C^{g(2)}$~\cite{DelDuca:2014cya} is entangled with the remainder $\cR^{(2,0)}_{4g}$ in the
non-logarithmic term of the two-loop amplitude,
\be
	{\rm Re} \left[ \cM^{(2,0)}_{4g}  \right] = 2 C^{g(2)}+\big(C^{g(1)}\big)^2 - \frac{\pi^2}{4}\big(\alpha^{(1)}\big)^2 + \cR^{(2,0)}_{4g}\,.\label{eq:twoloopN2LL} 
\ee 
Likewise, the three-loop Regge trajectory $\alpha^{(3)}$ is entangled with the remainder $\cR^{(3,1)}_{4g}$
in the single-logarithmic term of the three-loop amplitude,
\be
{\rm Re} \left[ \cM^{(3,1)}_{4g}  \right] = \alpha^{(3)}+ 2 \alpha^{(2)} C^{g(1)}+ \alpha^{(1)}\left( 2 C^{g(2)} + \big(C^{g(1)}\big)^2\right) - \frac{\pi^2}{4}\big(\alpha^{(1)}\big)^3 + \cR^{(3,1)}_{4g}\,.\label{eq:threeloopN2LL}
\ee 
Conversely, at NNLL accuracy the remainder does not contribute to the imaginary parts of the amplitude~\cite{DelDuca:2014cya}, which can then be written in terms of the same building blocks and provide an important consistency check of the factorisation.

Finally, at ${\rm N^3LL}$ accuracy, the three-loop impact factor $C^{g(3)}$ is entangled with the remainder $\cR^{(3,0)}_{4g}$ in the non-logarithmic term of the three-loop amplitude,
\be 
{\rm Re} \left[ \cM^{(3,0)}_{4g}  \right] = 2 C^{g(3)} + 2 C^{g(2)}C^{g(1)}-\frac{\pi^2}{2}\left( \alpha^{(2)}\alpha^{(1)}+\big(\alpha^{(1)}\big)^2\right)C^{g(1)} + {\rm Re} \left[ \cR^{(3,0)}_{4g} \right]\,.\label{eq:threeloopN3LL}
\ee 

Since the splitting in eq.\eqref{eq:factorisation&remainder} is not uniquely defined, fixing the remainder amounts to a scheme choice. One such scheme introduced in Ref.~\cite{DelDuca:2014cya} identifies the impact factor with the diagonal terms of the antisymmetric octet exchange and the factorisation-violating terms with the off-diagonal ones. The scheme introduced in Ref.~\cite{Falcioni:2021buo} on the other hand, dubbed the Regge-cut scheme (Rcs), identifies the non-planar part of the three-Reggeon exchange with the remainder $\cR^{[8_a]}_{4g}$ at three-loop accuracy, while the 
planar part is absorbed into the factorising part $\cM^{[8_a]\,\text{fact}}_{4g}$ of the amplitude.
The Regge-cut scheme is particularly useful, because it allows one to restrict the factorisation-violating contributions to the non-planar part of the theory, and thus to terms which are subleading in $N_c$. However, not all $N_c$-subleading terms belong to factorisation-violating contributions,
since $N_c$-subleading terms may be contained in the factorising part of the amplitude. Thus, by evaluating the planar part of the theory, we may fix uniquely the relevant BFKL building blocks, like the two-loop~\cite{DelDuca:2014cya} and three-loop impact factors and the three-loop Regge trajectory, though we can evaluate them only at leading colour (LC). To sum up, at three-loop accuracy we can set the remainder term in \eqn{eq:factorisation&remainder} to zero, and write
\be
\label{eq:planar}
	\left. \cM^{[8_a]}_{4g,\, \text{planar}}\right|_{\text{Rcs}} = \left. \cM^{[8_a]\,\text{fact}}_{4g}\right|_{\text{LC}} \,.
\ee
Furthermore, in the Regge-cut scheme the three-loop Regge trajectory in ${\cal N}=4$ SYM~\cite{Henn:2016jdu,Caron-Huot:2017fxr} agrees~\cite{Falcioni:2021buo} with the same quantity computed in planar ${\cal N}=4$ SYM~\cite{Drummond:2007aua,Naculich:2007ub,DelDuca:2008pj,DelDuca:2008jg}. Thus, in the Regge-cut scheme $N_c$-subleading terms are absent from the three-loop Regge trajectory in ${\cal N}=4$ SYM. That is in line with the expectation that the Regge trajectory is made of maximally non-Abelian colour structures only.

In this paper, we compute the two-loop -- and three-loop impact factors and the three-loop Regge trajectory of planar SU($N_c$) Yang-Mills theory, which through the Regge-cut scheme we may identify as the $N_c$-leading and $n_f$-independent contributions to the impact factors and the Regge trajectory in QCD. Further, it is conjectured that, like in ${\cal N}=4$ SYM, $N_c$-subleading terms are absent from the three-loop Regge trajectory. Thus, we interpret our result for the Regge trajectory of planar SU($N_c$) Yang-Mills theory as the QCD three-loop Regge trajectory at $n_f = 0$.

In sec.~\ref{sec:octet}, we detail how we project the antisymmetric octet out of the three-loop helicity amplitudes of planar $SU(N_c)$ Yang-Mills~\cite{Jin:2019nya}. In sec.~\ref{sec:buildingblocks}, we present the unrenormalised version of the two-loop impact factor, and of the three-loop Regge trajectory and impact factor. In sec.~\ref{sec:ifcomp}, we present the renormalised version of the same quantities, and we show that the infrared pole structure of the two-loop impact factor and of the three-loop Regge trajectory agrees with the
prediction of Ref.~\cite{DelDuca:2014cya}. In particular, we verify that the infrared poles of the renormalised three-loop Regge trajectory are given by the scale integral $K(\bar{\alpha}_s)$ over the cusp anomalous dimension~\cite{DelDuca:2014cya}, \eqn{eq:ren3regge}. If we give that for granted, the main result of this paper is to present for the first time the finite terms of the three-loop Regge trajectory of planar SU($N_c$) Yang-Mills theory, which we conjecture to coincide with the finite terms of the QCD three-loop Regge trajectory at $n_f = 0$. In app.~\ref{sec:appa}, we restore the dependence of the impact factors on a generic Regge factorisation scale.

\section{The Antisymmetric Octet at Three-Loop Order}
\label{sec:octet}

In Ref.~\cite{Jin:2019nya}, the planar $SU(N_c)$ Yang-Mills $gg \rightarrow gg$ amplitude was computed through three loops in all helicity configurations. In what follows we perform the expansion in the Regge limit detailed in the previous section in order to extract the three-loop Regge trajectory $\alpha^{(3)}_{\text{YM}}(\eps)$ and the three-loop impact factor $C_{\text{YM}}^{g(3)}$ in planar Yang-Mills theory. 

The planar amplitude $\cM_{4g,\,\text{planar}} $ contains only leading colour contributions, and thus it can be written in terms of colour-ordered amplitudes,
\begin{equation}
	\cM_{4g,\, \text{planar}} = \sum_{\sigma \in S_4/Z_4} \Tr(T^{a_{\sigma_1}}T^{a_{\sigma_2}}T^{a_{\sigma_3}}T^{a_{\sigma_4}})M_{4g,\, \text{planar}}(\sigma_1,\sigma_2,\sigma_3,\sigma_4)\,,
\end{equation}
where the $T^a$'s are the $SU(N_c)$ matrices in the fundamental representation. Since only leading colour structures are considered, the same colour decomposition is used at tree and at loop level, which motivates why we factored the tree amplitude out in \eqn{eq:OctetPerturbExpansion}.

Since our goal is to extract the building blocks appearing in the high-energy factorisation given in eq.~\eqref{eq:ReggeFactorisation}, we need to project the full amplitude onto the antisymmetric octet channel. This can be achieved by contracting it with the colour projector,
\begin{align}\nonumber
	P^{a_1 a_4}_{a_2 a_3}({8}_a) &=\frac{1}{N_c} f^{a_1 b a_4}f^{a_2 b a_3} \,.
\end{align}
which can be found e.g. in~\cite{DelDuca:1995hf}. We find that the antisymmetric octet can be obtained as
\begin{align}\label{eq:octetPerm}
	\cM_{4g}^{[8_a]}&= M(1,2,3,4) - M(1,3,2,4) - M(1,4,2,3) + M(1,4,3,2)\,,
\end{align}
where, for the sake of brevity, here and in the next equation we drop the $4g,\, \text{planar}$ subscript on the colour-ordered amplitudes.
Focusing now on the $(++--)$ helicity component, we can write
\begin{align}
\label{eq:octetPermHel}
	\cM_{++--}^{[8_a]} =&\, M(1^{+},2^{+},3^{-},4^{-}) - M(1^{+},3^{-},2^{+},4^{-})\nonumber\\
				       & \, - M(1^{+},4^{-},2^{+},3^{-}) + M(2^+,1^{+},4^{-},3^{-}) \,,
\end{align}
where we have used the cyclic invariance of colour-ordered amplitudes to match the helicity configuration of the last colour-ordered amplitude in \eqn{eq:octetPermHel} to the ones given in~\cite{Jin:2019nya}. 

We normalise by the tree-level amplitude $\cM_{++--}^{[8_a](0)}$ as in \eqn{eq:OctetPerturbExpansion}, 
and compute the BFKL building blocks by extracting them from the real part of the colour-ordered amplitude $M_{++--}$ given in~\cite{Jin:2019nya}. Further, we perform a comparison of the imaginary parts in the antisymmetric octet channel~\cite{DelDuca:2014cya}, which provides an extra layer of cross checks.

Finally, we comment on the regularisation scheme implicitly used to obtain these results. In~\cite{Jin:2019nya}, the computations are performed by making use of the conventional dimensional regularisation scheme (CDR), where both internal and external particles are presumed to live in a space of generic dimensionality. Helicity amplitudes are eventually obtained by setting the gauge invariant bases~\cite{Boels:2017gyc} in terms of which the result is written to explicit helicity configurations. That entails to project the amplitudes to two-dimensional external polarisation states. Accordingly,
the change of dimension of the polarisations of the external gluons implies a change in the regularisation scheme used in the computation from CDR to 't Hooft-Veltman (HV) scheme.
In fact, as detailed in sec.~\ref{sec:buildingblocks}, the building blocks we extract from this computation match established results which were computed in the HV scheme, in which the external gluons are restricted to four dimensions
and to two polarisation states.

\section{The BFKL Building Blocks}
\label{sec:buildingblocks}

In this section, we present the extracted building blocks up to three-loop order in pure Yang-Mills theory from the real part of the four-gluon amplitude~\cite{Jin:2019nya}. The presented building blocks are the coefficients of the Regge expansion in \eqn{eq:ReggeFactorisation} in terms of bare, rescaled couplings $\tilde{g}_s$ as in eqs.~\eqref{eq:permExpCGlu}~-~\eqref{eq:permExpTraj}.

At LL and NLL accuracy, using eqs.~(\ref{eq:oneloopLL}),~(\ref{eq:oneloopNLL}) and~(\ref{eq:twoloopNLL}), we obtain,
\begin{align}
	\alpha^{(1)}_{\text{YM}}(\epsilon) =& \, \frac{2}{\epsilon}\,,\\ 	
	C^{g(1)}_{\text{YM}}(\epsilon)=& - \frac{\gamma_K^{(1)}}{\epsilon^2} - \frac{\beta_0}{2\epsilon} + 2  \zeta_2 - \gamma_K^{(2)} - \text{AD} \, \epsilon- \bigg( \frac{1214}{81} - 3 \zeta_4 \bigg) \epsilon^2 \label{eq:oneloopimpa}\\
	&\,\, - \bigg( \frac{7288}{243} - \zeta_5 \bigg) \epsilon^3\,-\left( \frac{43736}{729}-3\zeta_6 \right)\epsilon^4 +  \cO(\epsilon^5)\,,\nonumber\\
		\alpha^{(2)}_{\text{YM}}(\epsilon) =& \, \frac{\beta_0}{\epsilon^2} + 2 \frac{\gamma_K^{(2)}}{\epsilon} 
	+ 2 \text{AD}
	+ \epsilon \left( \frac{2428}{81} - 66 \zeta_3 - 8 \zeta_4\right) \label{eq:2regge}\\ 
	&+ \epsilon^2 \left( \frac{14576}{243} -134 \zeta_3 - 99 \zeta_4 + 82 \zeta_5 + 36 \zeta_2 \zeta_3\right)+\mathcal{O}(\epsilon^3)\,,\nonumber
\end{align}	
where the first few orders in $\epsilon$ are given in terms of
the $n_f$-independent and $N_c$-rescaled beta function and cusp anomalous dimension,
\begin{align}\label{eq:betaAndCusp1}
	\beta_0 &=  \frac{11}{3}\,,\qquad\gamma_K^{(1)} = 2\,,\qquad\gamma_K^{(2)} = \frac{67}{18}-\zeta_2\,,\qquad \text{AD} =  \frac{202}{27}-\zeta_3\,,
\end{align}
and which match the established results from the literature, see e.g.~\cite{DelDuca:2017pmn}. Note that, for the sake of conciseness, in the impact factors (\ref{eq:oneloopimpa}),
(\ref{eq:cg2ym}) and (\ref{eq:cg3ym}), we set the Regge factorisation scale to $\tau = -t$. We restore the $\tau$ dependence in app.~\ref{sec:appa}.
The ${\cal O}(\eps)$ and ${\cal O}(\eps^2)$ of the two-loop Regge trajectory (\ref{eq:2regge}) agree with the $n_f$-independent terms of an unpublished evaluation\footnote{Bernhard Mistlberger, private communication.} of the QCD two-loop Regge trajectory, based on Ref.~\cite{Ahmed:2019qtg}.

At NNLL accuracy, we find for the two-loop gluon impact factor,
\begin{align} 
	C^{g(2)}_{\text{YM}}(\epsilon)=& \, \frac{\gamma_K^{(1)}}{\epsilon^4} + \frac{\beta_0}{2 \epsilon^3} 
	+ \bigg( \frac{49}{24}-5\zeta_2 \bigg) \frac1{\epsilon^2}-\bigg(\frac{199}{108} - \frac{55}{6}\zeta_2 +\zeta_3 \bigg)\frac1{\epsilon} \nn\\
        &-\bigg(\frac{16139}{648} -\frac{335}{18}\zeta_2 -\frac{121}{6}\zeta_3 +\frac{55}{4}\zeta_4 \bigg)\nonumber\\
	&- \bigg( \frac{8623}{72} - \frac{1037}{27}\zeta_2 - \frac{1865}{18}\zeta_3 - \frac{1199}{12}\zeta_4 - \zeta_2 \zeta_3 + 41\zeta_5 \bigg)\epsilon\nonumber\\ 
	&- \bigg( \frac{5 382 749}{11 664} - \frac{6232}{81}\zeta_2 - \frac{9410}{27}\zeta_3 - \frac{10687}{36}\zeta_4 + 209 \zeta_2 \zeta_3\nonumber\\
	&+ \frac{77}{6}\zeta_5 + \frac{95}{2}\zeta_3^2 + \frac{1695}{8}\zeta_6 \bigg) \epsilon^2 + \cO(\epsilon^3)\,,
	\label{eq:cg2ym}
\end{align}
where the maximal weight terms of \eqn{eq:cg2ym} agree
through ${\cal O}(\eps^2)$ with the two-loop impact factor~\cite{DelDuca:2008pj,DelDuca:2008jg} of planar ${\cal N}=4$ SYM. The ${\cal O}(\eps^0)$ terms of the QCD two-loop gluon impact factor have been previously presented in a different scheme~\cite{Caron-Huot:2017fxr}. The finite terms of \eqn{eq:cg2ym} may be understood as the
$N_c$-leading and $n_f$-independent finite terms of the QCD two-loop gluon impact factor in the Regge-cut scheme.

For the three-loop Regge trajectory, we obtain
\begin{align}
	\alpha^{(3)}_{\text{YM}}(\epsilon) =& \, \frac{242}{27}\frac1{\epsilon^3}  + \bigg( \frac{3254}{81} - \frac{88}{9} \zeta_2 \bigg) \frac1{\epsilon^2} + \bigg( \frac{11093}{81} - \frac{536}{27}\zeta_2-\frac{88}{9} \zeta_3  +\frac{44}{3}\zeta_4\bigg) \frac1{\epsilon} \nonumber\\
	\,\,&+ \frac{617525}{1458}-\frac{3196}{81}\zeta_2 -\frac{19732}{27}\zeta_3-\frac{253}{3}\zeta_4 +\frac{40}{3} \zeta_2 \zeta_3 + 16\zeta_5 +  \cO(\epsilon)\,.
	\label{eq:3ymtraj}
\end{align}
Truncating the three-loop trajectory up to terms only of maximal weight yields
\be
	\alpha^{(3)}_{\text{YM}}(\epsilon)\bigg|_{\text{max weight}} = \frac{44}{3}\frac{\zeta_4}{\epsilon} +\frac{40}{3} \zeta_2 \zeta_3 + 16\zeta_5 +  \cO(\epsilon)\,,
\ee
which matches the three-loop Regge trajectory in $\cN=4$ SYM~\cite{Drummond:2007aua,Naculich:2007ub,DelDuca:2008pj,DelDuca:2008jg,Henn:2016jdu,Caron-Huot:2017fxr,Falcioni:2021buo}.
Conjecturing that, like in ${\cal N}=4$ SYM, $N_c$-subleading terms are absent from the three-loop Regge trajectory, we may understand \eqn{eq:3ymtraj} as the pure gauge, or $n_f$-independent, part of the QCD unrenormalised three-loop Regge trajectory.

At N$^3$LL accuracy, using eq.~\eqref{eq:threeloopN3LL}, one may compute the three-loop gluon impact factor, 
\begin{align}
  	C^{g(3)}_{\text{YM}}(\epsilon)=&   -\frac{4}{3\epsilon^6} + \left(\frac{605}{162}+4 \zeta_2\right)\frac1{\epsilon^4}+\left(\frac{61525}{3888}-\frac{605}{27} \zeta _2\right) \frac1{\epsilon^3}\nonumber\\
  	&+\left(\frac{177121}{3888}-\frac{12419 }{648}\zeta _2-\frac{1100 }{27}\zeta _3+\frac{217 }{9}\zeta _4\right)\frac1{\epsilon^2}\nonumber\\
  	&+\left(\frac{2489669}{34992}+\frac{28663}{972} \zeta _2-\frac{75265}{648} \zeta _3-\frac{18865}{72} \zeta _4-\frac{22}{9}  \zeta _3 \zeta _2+\frac{224 }{3}\zeta _5\right)\frac1{\epsilon}\nonumber\\
  	&-\frac{10881647}{52488}+\frac{1510567}{5832} \zeta _2+\frac{29113}{108} \zeta _3-\frac{18895}{216} \zeta _4-\frac{1639}{9} \zeta _5\nonumber\\
  	&+\frac{27269}{54} \zeta _2 \zeta _3+\frac{796}{9} \zeta _3^2+\frac{211861}{432} \zeta _6 + \cO(\epsilon)\,,
  	\label{eq:cg3ym}
\end{align}
where again the maximal weight terms of \eqn{eq:cg3ym} agree
through ${\cal O}(\eps^0)$ with the three-loop impact factor~\cite{DelDuca:2008pj,DelDuca:2008jg} of planar ${\cal N}=4$ SYM. As for the two-loop impact factor (\ref{eq:cg2ym}),
\eqn{eq:cg3ym} may be interpreted as the
$N_c$-leading and $n_f$-independent part of the QCD three-loop gluon impact factor in the Regge-cut scheme.

\section{Comparison to Infrared Structure}
\label{sec:ifcomp}

In~\cite{DelDuca:2014cya} a comparison between the infrared factorisation and the Regge limit of $2\to 2$ scattering amplitudes led to results on the infrared pole structure of the building blocks 
we have presented in sec.~\ref{sec:buildingblocks} up to NNLL  accuracy. It is however not straightforward to compare to this analysis, since it was done for renormalised amplitudes so as to disentangle the ultraviolet and infrared behaviour. On the other hand, we have expressed our building blocks as coefficients of the bare, rescaled couplings $\tilde{g}_s$ defined in eq.~\eqref{eq:gtildedef}. The two versions of the building blocks can however easily be translated from one convention to the other by renormalising the generic Regge-factorised amplitude given in eq.~\eqref{eq:ReggeLogExpansion} and comparing coefficients with its equivalent expansion in terms of the renormalised coupling $\bar\alpha_s$, implicitly defined by
\begin{equation}
	\alpha_s = \bar\alpha_s Z_\alpha(\bar\alpha_s,\epsilon) \,,
\end{equation}
with a $Z$-factor given by
\begin{align}
    Z_\alpha(\bar\alpha_s,\epsilon) &= 1 - \bar\alpha_s \frac{\beta_0 }{\epsilon} +\bar\alpha_s^2 \left(\frac{\beta_0^2}{\epsilon^2}-\frac{\beta_1}{2 \epsilon}\right) + \bar\alpha_s^3 \left(-\frac{\beta_0^3}{\epsilon^3}+\frac{7 \beta_0 \beta_1}{6 \epsilon^2}-\frac{\beta_2}{3 \epsilon}\right) \,,
\end{align}
with $\beta_0$ in eq.~\eqref{eq:betaAndCusp1}, and where the
$n_f$-independent and $N_c$-rescaled higher orders of the beta function are
\begin{equation}\label{eq:higherorderbeta}
	\beta_1 = \frac{34}{3}\,, \qquad \beta_2=\frac{2857}{54} \,.
\end{equation}
For the perturbative expansion (\ref{eq:ReggeLogExpansion}), 
this results in
\begin{align} \label{eq:pre_bareToRen}
	\cM^{[8_a]}_{4g} &= \cM^{[8_a](0)}_{4g}\left(1 + \sum_{L=1}^{\infty} N_c^L \left(   Z_\alpha(\bar\alpha_s,\epsilon)\frac{\kappa_\Gamma}{S_\epsilon}\frac{\bar\alpha_s}{4 \pi} \right)^{L} \sum_{i=0}^{L}  \cM^{(L,i)}_{4g} \log^i\left( \frac{s}{\tau} \right)  \right) \\ \nonumber
				&= \cM^{[8_a](0)}_{4g}\left(1 + \sum_{L=1}^{\infty} N_c^L \left(\frac{\bar\alpha_s}{4 \pi}\right)^{L} \sum_{i=0}^{L}  \cM^{(L,i)}_{4g,\,\text{R}} \log^i\left( \frac{s}{\tau} \right)  \right) \,,
\end{align}
where we set $\mu^2 = -t$ and where the coefficients $\cM^{(L,i)}_{4g}$ are expressed in terms of the building blocks as introduced in eqs.~\eqref{eq:oneloopLL}~-~\eqref{eq:threeloopN2LL}, while the coefficients $\cM^{(L,i)}_{4g,\,\text{R}}$ are expressed in terms of renormalised building blocks,
\begin{align}
    \cM^{(L,i)}_{4g,\,\text{R}} =  \cM^{(L,i)}_{4g} \bigg|_{
    \substack{\alpha^{(L)} \leftrightarrow \, \alpha^{(L)}_{\text{R}}\\ C^{g(L)} \leftrightarrow \, C^{g(L)}_{\text{R}}}
    } \,.
\end{align}
In this way, we obtain a direct translation from the bare building blocks $\alpha^{(L)}$ and $C^{g(L)}$ to the renormalised building blocks $\alpha^{(L)}_\text{R}$ and $C^{g(L)}_\text{R}$, which is given by
\begin{align}\label{eq:bareToRen_al1}
    \alpha^{(1)}_{\text{R}} &= \alpha^{(1)} \frac{\kappa _{\Gamma }}{S_\epsilon} \,,\\
    \alpha^{(2)}_{\text{R}} &= \alpha^{(2)} \left(\frac{\kappa _{\Gamma }}{S_\epsilon}\right)^2-\left(\alpha^{(1)} \beta _0 \frac{\kappa _{\Gamma }}{S_\epsilon}\right) \frac{1}{ \epsilon}  \,,\\ 
    \alpha^{(3)}_{\text{R}} &= \alpha^{(3)} \left(\frac{\kappa _{\Gamma }}{S_\epsilon}\right)^3 - \left( 2 \alpha^{(2)} \beta _0 \left(\frac{\kappa _{\Gamma }}{S_\epsilon}\right)^2 +  \frac{1}{2}\alpha^{(1)}\beta_1  \frac{\kappa _{\Gamma }}{S_\epsilon} \right)\frac{1}{\epsilon} + \left(\alpha^{(1)} \beta _0^2 \frac{\kappa _{\Gamma }}{S_\epsilon}\right)\frac{1}{ \epsilon^2}\,, \\ 
    C^{g(1)}_{\text{R}} &= C^{g(1)} \frac{\kappa _{\Gamma }}{S_\epsilon}-\frac{\beta _0}{2  \epsilon}\,, \\
    C^{g(2)}_{\text{R}} &= C^{g(2)} \left(\frac{\kappa _{\Gamma }}{S_\epsilon}\right)^2 - \left(\frac{6}{4}\beta _0 C^{g(1)} \frac{\kappa _{\Gamma }}{S_\epsilon}  + \frac{\beta _1}{4} \right)\frac{1}{ \epsilon}+\frac{3 \beta _0^2}{8  \epsilon^2}\,\label{eq:bareToRen_cGlu2},\\ 
    C^{g(3)}_{\text{R}} &= C^{g(3)} \left(\frac{\kappa _{\Gamma }}{S_
    \epsilon}\right)^3 - \left( \frac{5}{2} \beta _0 C^{g(2)} \left(\frac{\kappa _{\Gamma }}{S_{\epsilon}}\right)^2 + \frac{3}{4} \beta _1 C^{g(1)} \frac{\kappa _{\Gamma }}{S_{\epsilon}} + \frac{\beta _2}{6}  \right)\frac{1}{ \epsilon}\nonumber \\ \label{eq:bareToRen_cGlu3}
    &~~~+  \left( \frac{15}{8} \beta _0^2 C^{g(1)} \frac{\kappa _{\Gamma }}{S_\epsilon} + \frac{11 \beta _1 \beta _0}{24}  \right)\frac{1}{ \epsilon^2} -\frac{5 \beta _0^3}{16  \epsilon^3}\,.
\end{align}

After inserting the bare building blocks into eqs.~\eqref{eq:bareToRen_al1}~-~\eqref{eq:bareToRen_cGlu2} and expanding the right-hand side in $\epsilon$ to the appropriate orders, we find the following explicit results for the renormalised BFKL building blocks,
\begin{align}
	\alpha^{(1)}_{\text{R,YM}} =&\frac{\gamma_K^{(1)}}{\epsilon } + \cO\left(\epsilon^0\right)\,,\\
	\alpha^{(2)}_{\text{R,YM}} =&-\frac{\beta _0 \gamma_K^{(1)}}{2 \epsilon ^2}+\frac{2 \gamma_K^{(2)}}{\epsilon } + \cO\left(\epsilon^0\right) \,,\\
	\alpha^{(3)}_{\text{R,YM}} =& \frac{\beta _0^2 \gamma_K^{(1)}}{3 \epsilon ^3}- \left(\frac{1}{3} \beta _1 \gamma_K^{(1)}+\frac{4}{3}\beta _0 \gamma_K^{(2)}\right)\frac{1}{\epsilon ^2} +\frac{16 \gamma_K^{(3)}}{3 \epsilon } + \cO\left(\epsilon^0\right) \,,\\
	C^{g(1)}_{\text{R,YM}}=& -\frac{2}{\epsilon ^2} - \frac{\beta _0}{\epsilon }  + \cO\left(\epsilon^0\right) \,, \\
	C^{g(2)}_{\text{R,YM}}=& \frac{2}{\epsilon ^4} + \frac{7 \beta _0}{2 \epsilon ^3}+ \left( \frac{103}{6}-7\, \zeta_2 \right)\frac{1}{\epsilon ^2} + \left(\frac{853}{54} -\frac{44}{3}   \zeta_2 - \frac{31}{3} \zeta_3 \right)\frac{1}{ \epsilon } + \cO\left(\epsilon^0\right) \,,\\
	C^{g(3)}_{\text{R,YM}}=& - \frac{4}{3\epsilon^6}-\frac{5\beta_0}{\epsilon^5}-\left( \frac{10285}{162} - 6\zeta_2 \right)\frac{1}{\epsilon^4}-\left( \frac{46181}{972} - \frac{2255}{54}\zeta_2 -\frac{28}{3}\zeta_3 \right)\frac{1}{\epsilon^3}\nonumber\\
	&+\left( \frac{1577}{486} + \frac{5731}{648}\zeta_2 + \frac{2915}{54}\zeta_3 + \frac{661}{36}\zeta_4 \right)\frac{1}{\epsilon^2}+\left( \frac{2338843}{17496} - \frac{76315}{486}\zeta_2\right.\nonumber\\
	& \left.- \frac{11435}{108}\zeta_3 - \frac{1265}{18}\zeta_4 - \frac{400}{9}\zeta_2\zeta_3 + \frac{1492}{15}\zeta_5 \right)\frac{1}{\epsilon}+\cO(\epsilon^0)
\end{align}
where up to two-loop order the coefficients of the cusp anomalous dimension and beta function are given in eqs.~\eqref{eq:betaAndCusp1} and~\eqref{eq:higherorderbeta}, and the three-loop cusp anomalous dimension is
\begin{equation}
    \gamma_K^{(3)}= \frac{245}{48} - \frac{67}{18}\zeta_2 + \frac{11}{12}\zeta_3 +\frac{11}{4}\zeta_4\,.
\end{equation}
Note that, up to a difference in normalisation\footnote{Which is entirely due to expanding the amplitude in $\alpha_s/(4\pi)$ as opposed to $\alpha_s/\pi$.}, we reproduce the results for the one- and two-loop building blocks presented in~\cite{DelDuca:2014cya}. In particular, the poles of the two-loop impact factor agree with the pure gauge, or $n_f$-independent, poles of the QCD two-loop impact factor~\cite{DelDuca:2014cya}.

Further, at each perturbative order and up to three loops,
the poles of the gluon Regge trajectory are  entirely fixed by the scale integral $K(\bar{\alpha}_s)$ over the cusp anomalous dimension, whose expansion to three-loop order is 
\begin{align}
    K(\bar{\alpha}_s) =& \frac{\bar{\alpha}_s}{4 \pi}\frac{{\gamma}_{K}^{(1)}}{\epsilon}+\left(\frac{\bar{\alpha}_s}{4 \pi}\right)^2 \left( \frac{2 \gamma_K^{(2)}}{\epsilon } -\frac{\beta _0 \gamma_K^{(1)}}{2 \epsilon ^2}\right)\nonumber\\
    &+\left(\frac{\bar{\alpha}_s}{4 \pi}\right)^3\left( \frac{16 }{3 \epsilon }\gamma_K^{(3)} -\frac{\beta _1 \gamma_K^{(1)}+4\beta _0 \gamma_K^{(2)}}{3 \epsilon ^2}+\frac{\beta _0^2 \gamma_K^{(1)}}{3 \epsilon ^3}\right) + \cO(\bar{\alpha}_s^4) \,,
\end{align}
as found in~\cite{DelDuca:2014cya}\footnote{The term $\sim \beta_1$ in the $1/\epsilon^2$ pole differs by a factor of 4 from the result in~\cite{DelDuca:2014cya} since the coefficients of the beta function $\beta_i$ in this paper are extracted from an expansion in $\alpha_s/(4\pi)$. By contrast, our normalisation for the coefficients of the cusp anomalous dimension $\gamma_K^{(i)}$ is such that they are extracted from the all-order cusp through an expansion in $\alpha_s/\pi$. }. As expected~\cite{DelDuca:2014cya,Falcioni:2021buo}, 
we find that at three-loop order all the poles of the renormalised Regge trajectory are captured by $K(\bar{\alpha}_s)$, i.e.
\be
\alpha_{\text{R,YM}} = K(\bar{\alpha}_s) + {\cal O}(\eps^0) 
\label{eq:ren3regge}
\ee 
holds at three-loop order.

\section{Conclusions}
\label{sec:concl}

In this paper we have computed the two-loop impact factor (\ref{eq:cg2ym}), the three-loop Regge trajectory (\ref{eq:3ymtraj}) and the three-loop impact factor (\ref{eq:cg3ym}) in pure Yang-Mills theory at leading colour through $\cO\left(\epsilon^0\right)$. 
Their maximum-weight components match the three-loop Regge trajectory~\cite{Drummond:2007aua,Naculich:2007ub,DelDuca:2008pj,DelDuca:2008jg,Henn:2016jdu,Caron-Huot:2017fxr,Falcioni:2021buo} and the two-loop and three-loop impact factors~\cite{DelDuca:2008pj,DelDuca:2008jg} of $\cN = 4$ SYM.
Further, the two-loop impact factor and the three-loop Regge trajectory display the infrared behaviour predicted in~\cite{DelDuca:2014cya}.

We understand \eqns{eq:cg2ym}{eq:cg3ym} to be the $N_c$-leading and $n_f$-independent parts of the QCD two-loop and three-loop gluon impact factors, respectively.
Conjecturing that, like in ${\cal N}=4$ SYM, $N_c$-subleading terms are absent from the three-loop Regge trajectory, we understand \eqn{eq:3ymtraj} to be the pure gauge, or $n_f$-independent, part of the QCD three-loop gluon Regge trajectory. This is an essential ingredient in the computation of the BFKL equation at next-to-next-to leading logarithmic accuracy.

We present both the bare and the renormalised Regge trajectories and impact factors, and provide the dependence of the impact factor on the Regge factorisation scale for the sake of future matching with higher-multiplicity contributions.

\section*{Acknowledgements}

We thank Giulio Falcioni for very helpful discussions and Bernhard Mistlberger for providing an evaluation of the QCD two-loop Regge trajectory to ${\cal O}(\eps^2)$.
RM was supported by the ERC starting grant 757978 and by the research grant 00025445 from Villum Fonden. 
BV was supported by the European Research Council under ERC-STG-804286 UNISCAMP and by the Knut and Alice Wallenberg Foundation under grant KAW2018.0162.

\appendix
\section{Gluon impact factors with full $\tau$ dependence}
\label{sec:appa}

In this section we outline how to reinstate the dependence on the factorisation scale $\tau$ from the results presented in sec.~\ref{sec:buildingblocks}, which were given for $\tau = -t$. The explicit dependence on $\tau$ will be useful when combining results at different loop orders with different numbers of external legs when applying the BFKL formalism. As $\tau$ enters only through the ratios $\pm s/\tau$ in eq.~\eqref{eq:ReggeFactorisation}, it suffices to simply rewrite the large logarithms used for the extraction at $\tau = -t$ in sec.~\ref{sec:buildingblocks} as
\begin{equation}
    \log \left( \frac{s}{-t} \right) = \log \left( \frac{s}{\tau} \right) - \log \left( \frac{-t}{\tau} \right)\,,
\end{equation}
and to compare the coefficients of the logarithms $\log (-t/\tau)$ as we have done before. We find the following relations between the $\tau$-independent building blocks $C^{g}$ and the $\tau$-dependent building blocks $C^{g}_{\tau}$,
\begin{align}
C^{g(1)}_{\tau} =&\, C^{g(1)}-\frac{1}{2} \alpha^{(1)} \log \left(\frac{ -t}{\tau }\right)\,, \label{eq:if1tau}\\
 C^{g(2)}_{\tau} =& \, C^{g(2)} - \frac{1}{2} (\alpha^{(1)} C^{g(1)}+\alpha^{(2)}) \log \left(\frac{ -t}{\tau }\right)+\frac{1}{8} (\alpha^{(1)})^2 \log ^2\left(\frac{ -t}{\tau }\right)\,, \label{eq:if2tau}\\ 
 C^{g(3)}_{\tau} =& \, C^{g(3)} - \frac{1}{2} (\alpha^{(2)} C^{g(1)}+\alpha^{(1)} C^{g(2)}+\alpha^{(3)}) \log \left(\frac{ -t}{\tau }\right) \nonumber\\
 &+ \frac{1}{8} \alpha^{(1)} (\alpha^{(1)} C^{g(1)}+2 \alpha^{(2)}) \log ^2\left(\frac{ -t}{\tau }\right) - \frac{1}{48} (\alpha^{(1)})^3 \log ^3\left(\frac{ -t}{\tau }\right) \,,
\label{eq:if3tau}
\end{align}
and analogously for the renormalised case. The maximum-weight components of eqs.~(\ref{eq:if1tau})-(\ref{eq:if3tau}) match the $\tau$-dependent impact factors~\cite{DelDuca:2008pj,DelDuca:2008jg} of $\cN = 4$ SYM.


\bibliography{ref}

\end{document}